\documentclass[conference]{IEEEtran}
\IEEEoverridecommandlockouts
\usepackage{cite}
\usepackage{amsmath,amssymb,amsfonts}
\usepackage{algorithmic}
\usepackage{graphicx}
\usepackage{textcomp}
\usepackage{xcolor}
\usepackage{subcaption}
\usepackage{fancyhdr}
\usepackage[binary-units=true]{siunitx}
\usepackage[a4paper, total={184mm,239mm}]{geometry}
\def\BibTeX{{\rm B\kern-.05em{\sc i\kern-.025em b}\kern-.08em
    T\kern-.1667em\lower.7ex\hbox{E}\kern-.125emX}}
\DeclareMathOperator*{\softmax}{SoftMax}

\usepackage{acronym}
\acrodef{EMG}[EMG]{electromyography}
\acrodef{sEMG}[sEMG]{Surface electromyography}
\acrodef{HMI}[HMI]{Human-Machine Interface}
\acrodef{DL}[DL]{Deep Learning}
\acrodef{ML}[ML]{Machine Learning}
\acrodef{SoA}[SoA]{State-of-the-Art}
\acrodef{AP}[AP]{Action Potential}
\acrodef{MUAP}[MUAP]{Motor Unit Action Potential}
\acrodef{MUAPT}[MUAPT]{MUAP Train}
\acrodef{PLI}[PLI]{Power Line Intereference}
\acrodef{MAV}[MAV]{Mean Absolute Value}
\acrodef{WL}[WL]{Waveform Length}
\acrodef{RMS}[RMS]{Root Mean Square}
\acrodef{$k$-NN}[$k$-NN]{$k$-Nearest Neighbors} 
\acrodef{SVM}[SVM]{Support Vector Machine} 
\acrodef{RBF}[RBF]{Radial Basis Function}
\acrodef{RF}[RF]{Random Forest}
\acrodef{HDC}[HDC]{Hyper-Dimensional Computing}
\acrodef{MLP}[MLP]{Multi-Layer Perceptron}
\acrodef{CNN}[CNN]{Convolutional Neural Network}
\acrodef{RNN}[RNN]{Recurrent Neural Network}
\acrodef{LSTM}[LSTM]{Long Short-Term Memory}
\acrodef{TCN}[TCN]{Temporal Convolutional Network}
\acrodef{ReLU}[ReLU]{Rectified Linear Unit}
\acrodef{BN}[BN]{Batch-Normalization}
\acrodef{FC}[FC]{Fully Connected}
\acrodef{MCU}[MCU]{microcontroller unit}
\acrodef{SoC}[SoC]{System on Chip}

\acrodef{MAC}[MAC]{multiply-and-accumulate}
\acrodef{MAE}[MAE]{Mean Absolute Error}
\acrodef{DoF}[DoF]{Degree of Freedom}
\acrodef{DoA}[DoA]{Degree of Actuation}
\acrodef{EMA}[EMA]{Exponential Moving Average}
\acrodef{}[]{}
\acrodef{}[]{}

\renewcommand{\baselinestretch}{0.975}
\begin{document}
\bstctlcite{IEEEexample:BSTcontrol}

\title{Bioformers: Embedding Transformers for Ultra-Low Power sEMG-based Gesture Recognition
\thanks{This   work   was   supported   in   part   by  the EU Grant Bonsapp  (g.a. no. 101015848).}
}

\author{\IEEEauthorblockN{Alessio Burrello\IEEEauthorrefmark{1}, Francesco Bianco Morghet\IEEEauthorrefmark{2}, Moritz Scherer\IEEEauthorrefmark{3}, Simone Benatti\IEEEauthorrefmark{4}, \\Luca Benini\IEEEauthorrefmark{1}\IEEEauthorrefmark{3}, Enrico Macii\IEEEauthorrefmark{5}, Massimo Poncino\IEEEauthorrefmark{2}, Daniele Jahier Pagliari\IEEEauthorrefmark{2}}

\IEEEauthorblockA{
\IEEEauthorrefmark{1} DEI, Università di Bologna, Bologna, Italy \\
\IEEEauthorrefmark{2} Department of Control and Computer Engineering, Politecnico di Torino, Turin, Italy \\
\IEEEauthorrefmark{3} Integrated Systems Laboratory, ETH Zurich, Switzerland \\
\IEEEauthorrefmark{4} Department of Sciences and Methods for Engineering, University of Modena and Reggio Emilia, Italy\\
\IEEEauthorrefmark{5}Interuniversity Department of Regional and Urban Studies and Planning, Politecnico di Torino, Turin, Italy}

\IEEEauthorblockA{Emails: name.surname@unibo.it, name.surname@polito.it, simone.benatti@unimore.it, scheremo@iis.ethz.ch}
}

\maketitle

\begin{abstract}
Human-machine interaction is gaining traction in rehabilitation tasks, such as controlling prosthetic hands or robotic arms.
Gesture recognition exploiting surface electromyographic (sEMG) signals is one of the most promising approaches, given that sEMG signal acquisition is non-invasive and is directly related to muscle contraction.
However, the analysis of these signals still presents many challenges since similar gestures result in similar muscle contractions. Thus the resulting signal shapes are almost identical, leading to low classification accuracy.
To tackle this challenge, complex neural networks are employed, which require large memory footprints, consume relatively high energy and limit the maximum battery life of devices used for classification.
This work addresses this problem with the introduction of the Bioformers. This new family of ultra-small attention-based architectures approaches state-of-the-art performance while reducing the number of parameters and operations of 4.9$\times$. 
Additionally, by introducing a new inter-subjects pre-training, we improve the accuracy of our best Bioformer by 3.39\%, matching state-of-the-art accuracy without any additional inference cost.

Deploying our best performing Bioformer on a Parallel, Ultra-Low Power (PULP) microcontroller unit (MCU), the GreenWaves GAP8, we achieve an inference latency and energy of 2.72~ms and 0.14~mJ, respectively, 8.0$\times$ lower than the previous state-of-the-art neural network, while occupying just 94.2 kB of memory.
\end{abstract}

\section{Introduction}
\thispagestyle{fancy}
\fancyhf{}
\chead{Published as a conference paper at the IEEE 2022 DATE Conference}
In the last few years, thanks to the availability of increasingly powerful yet energy-efficient devices, there has been a consistent trend towards moving computations to the edge, therefore eliminating the need of relying on a centralized computational unit~\cite{madakam2015internet}. 
One of the fields that benefits the most from this new paradigm is personalized healthcare. Many applications, such as heart rate (HR) monitoring \cite{risso2021robust}, have been moved to wearable devices, reducing overall energy consumption thanks to the drastic reduction of raw data communication. Further, thanks to edge computing, new applications such as closed-loops brain stimulations, which require reliable, low-latency processing, become viable~\cite{sun2008responsive}.

Human Machine Interfaces (HMIs) can be extremely useful in personalized healthcare, enabling new types of interactions for impaired patients, for example, through hand gestures~\cite{meattini2018semg}.
A reliable approach for gestures recognition exploits surface electromyographic signal (sEMG), which has been demonstrated to be strongly correlated to the different arms positions and muscle contractions.
In this approach, a pre-defined set of gestures is selected, and training data are collected letting subjects perform the different gestures following a pre-defined pattern during the recording sessions. Then, a classification algorithm is trained to distinguish the different hand gestures based on the sEMG signals gathered. 
Currently, \ac{DL} algorithms~\cite{hu2018novel, Tsinganos2018, tsinganos2019improved} are the state-of-the-art for this task.

However, most of these advanced algorithms are too computationally expensive to be deployed on memory-constrained edge devices, occupying a too large memory footprint or necessitating a high amount of operations, hampering the battery lifetime of edge devices~\cite{hu2018novel, tsinganos2019improved, Betthauser2019}.
For this reason, several recent works focus on designing lightweight yet accurate DL models that can be deployed on edge devices. These works typically target platforms with less than 1MB memory and with a power envelope in the order of tens of mW \cite{gap8}.
Such tight constraints are typically met leveraging optimizations at different levels, which affect the model architecture (e.g., Neural Architecture Search and pruning \cite{risso2021robust}) and the bit-width used for storing and processing its parameters and intermediate outputs (quantization). These combined optimizations have recently enabled the deployment of several highly accurate models for healthcare applications at the edge \cite{risso2021robust}.

However, there is a clear pace difference in terms of DL model innovation between the cloud and the edge. On large scale classification problems, such as natural language processing (NLP) and, more recently, computer vision (CV), Transformer networks are quickly becoming state-of-the-art, outclassing all competitors in terms of accuracy. These results are obtained thanks to models such as BERT \cite{devlin2018bert}, GPT-3 \cite{brown2020language} and the VisionTransformer (ViT) \cite{dosovitskiy2020image}, which include hundreds of millions, or even billions of parameters. In contrast, most successful examples of DL deployment at the edge are based on Convolutional Neural Networks (CNNs) \cite{howard2019searching, burrello2020predicting}.

In this work, we aim to demonstrate that Transformers are also suitable for smaller-scale problems and that they can achieve state-of-the-art performance in a TinyML gesture recognition scenario with lower computational complexity compared to CNNs.
Furthermore, inspired by the large-scale pre-training typically applied to transformers for CV and NLP, we design a new pre-training strategy for sEMG-based gesture recognition.
Specifically, the contribution of our work is threefold:
\begin{itemize}
    \item We introduce a novel DL architecture for gesture recognition, the \textit{Bioformer}, which exploits the attention mechanism to reduce computational complexity while achieving state-of-the-art gestures recognition results.
    \item We introduce an inter-subject pre-training step to improve deep learning architectures' representation capability in gesture recognition. While this task is typically patient-specific, we show that employing data from other subjects aids the network to extract more significant and generalizable features.
    \item We demonstrate the advantages of using an initial 1D-convolutional layer to aggregate raw signals in a series of projections to feed the transformer network. Specifically, this layer increases accuracy and reduces complexity simultaneously.
\end{itemize}
Testing our architecture on the Ninapro DB6 dataset, which includes eight grasp gestures from 10 subjects, our best network achieves 62.34\% accuracy, further improved to 65.73\% thanks to the inter-subject pre-training.
Quantized to 8bits, it occupies as little as 94.2 kB, which is 4.9$\times$ lower than previous state-of-the-art CNN, TEMPONet \cite{zanghieri2019robust, zanghieri2020temporal}, achieving 65.0\% on the same task.
Deployed on the GAP8 multi-core MCU, the same Bioformer only consumes 0.139 mJ per inference, being 8.0$\times$ more efficient than TEMPONet.

\section{Background \& Related Work}
\label{sec:background_related}
\subsection{Surface Electromyographic Signal}
EMG signals \cite{de1997use} originate from the electrical activity that occurs during a muscular contraction, ranging from $\SI{10}{\micro\volt}$ to $\SI{1}{mV}$ with a bandwidth of $\sim \SI{2}{kHz}$ for standard applications, even though it is possible to acquire EMG data up to $\sim \SI{10}{kHz}$ in Motor Unit Action Potential Analysis. EMG activity is acquired by conductive plates (i.e. electrodes) placed on the skin surface that collect the underlying electrical muscular activity. A major issue of signal acquisition is related to the skin-electrode interface, which is prone to high variability and can degrade signal quality. Also, electrode re-positioning and user adaptation~\cite{zanghieri2019robust}, as well as motion artefacts and floating ground noise, represent major causes of signal degradation and variability.

\subsection{Related Work}
In the last few years, sEMG-based hand gesture recognition has gained traction in academia and commercial applications. Early approaches rely on conventional ML methods, such as Support Vector Machine (SVM), Random Forests (RFs), Linear Discriminant Analysis (LDA) or shallow artificial neural networks (ANN)~\cite{kaufmann2010fluctuating, Atzori_DB1_2014, Milosevic2017}. 
Even though SoA recognition accuracy is above 90\%, most experiments are limited to a single-session setup, failing to cope with the inter-session accuracy drop observed when gesture inference is made on sessions never seen during training \cite{Atzori_DB1_2014}. 
In fact, one of the major challenges in sEMG-based gesture recognition is managing the performance drop due to electrode multiday donning-doffing~\cite{PalermoNinaProDB6, Milosevic2017}. 
To tackle this issue, some approaches rely on extending the training datasets or increasing the electrode count, obtaining larger sets of features that improve algorithm convergence ~\cite{kaufmann2010fluctuating}. 
However, performance drops are still consistent, and the lack of generalization hinders the deployment of these solutions in real-world scenarios. 
Multi-session training is a promising strategy explored by virtue of the release of several multi-session sEMG datasets, such as the Non-Invasive Adaptive hand Prosthetics Database 6 (NinaPro DB6, 10 sessions, 8 classes)~\cite{PalermoNinaProDB6} and the University of Bologna - INAIL (Unibo-INAIL) database (8 days $\times$ 4 arm postures, 6 gestures)~\cite{Milosevic2017}.
These works show increased accuracy recognition over time but strongly rely on domain-specific knowledge and hand-crafted features. 
To increase model robustness and eliminate the need for hand-crafted feature extraction, DL-based models represent a viable solution, also prompted by the availability of relatively large public multi-session datasets.
For instance,~\cite{Atzori2016_DL} propose Convolutional Neural Networks (CNN), which outperform a SVM in classification accuracy across several subjects on a public dataset.
More recently, Temporal Convolutional Networks (TCNs), variants of a CNN widely used for time-series analysis ~\cite{tsinganos2019improved,Betthauser2019}, are gaining traction also in sEMG signal processing, showing high accuracy in multi-session problems. 
Even though DL-based approaches tackle the EMG variability problem successfully, they are based on large models. Hence, their deployment on real-time, resource-constrained edge platforms, such as wristbands or smartwatches, is still a non-trivial task. 

\subsection{Attention \& Transformers}
\label{subsec:attention}
\begin{figure}[t]
  \centering
\includegraphics[width=1.0\linewidth]{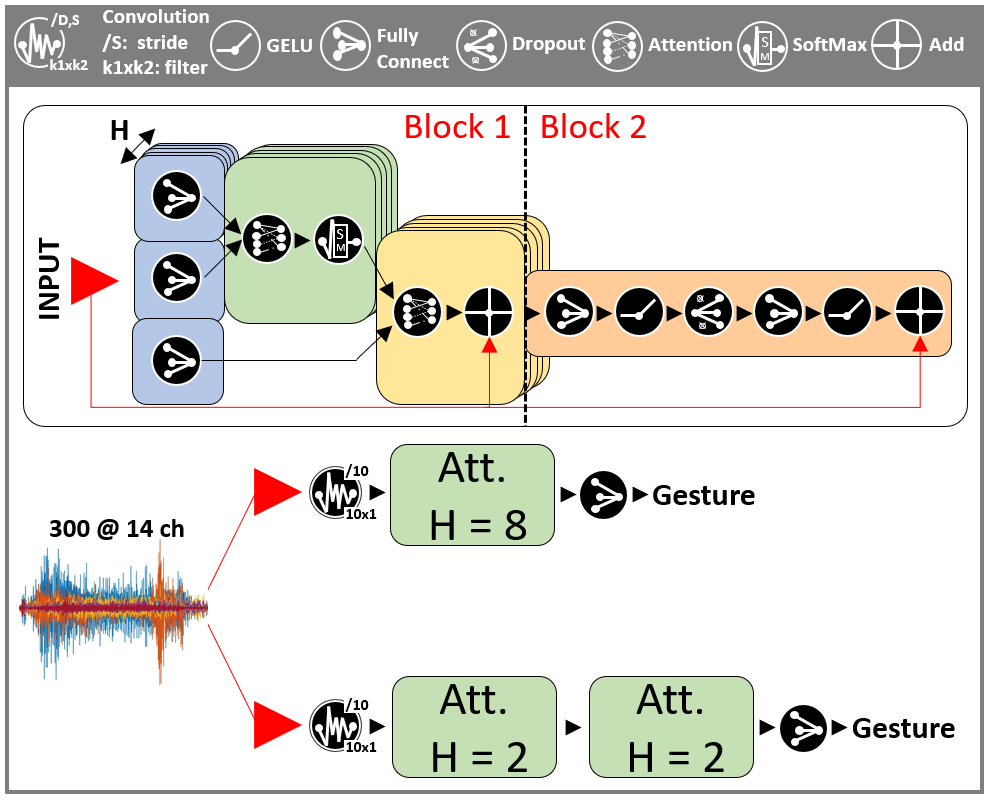}
  \caption{In the upper part, the basic MHSA layer used inside our architectures. In the lower part, the two Bioformers architectures that we propose as benchmarks.}
  \label{fig:bioformers}
  \vspace{-0.6cm}
\end{figure}
In 2017, \cite{vaswani2017attention} demonstrated the possibility to exploit a neural network solely based on the \textit{attention} mechanism to improve the performance of different language modelling tasks.
In DL, the concept of attention refers to layers that generate input-dependent synaptic weights, resulting in a variable relation among inputs that depends on the relative importance from the point of view of the target task (i.e., paying "more attention" to the most important inputs).
In \cite{vaswani2017attention}, the authors exploited in particular the so-called Multi-Heads Self-Attention (MHSA) blocks that analyze the relationship of different parts of the input data among themselves.
Similarly, in our work, we employ the MHSA as the basic building block of our architectures.

Given a tensor $\mathbf{X}$, with dimension $S \times C$, where $S$ is the \textit{input (or sequence) length} and $C$ the number of \textit{channels}, the MHSA produces an output of the same shape $S \times C$.
It comprises two main blocks, as shown on the top of Fig. \ref{fig:bioformers}.
The first one uses a set of parallel and independent heads, $H$, all of which perform a series of three operations on the input data (Blue, green and yellow rectangles in Fig.\ref{fig:bioformers}).
The first operation projects the sequence $\mathbf{X}$ into three separate \textit{projection spaces} each of size $P$, using three trainable linear layers. These projections are called \textit{queries} $\mathbf{Q}$, \textit{keys} $\mathbf{K}$ and \textit{values} $\mathbf{V}$, and are computed as:
\begin{equation}
    \mathbf{Q} = \mathbf{X}\mathbf{W}_\text{query} \qquad
    \mathbf{K} = \mathbf{X}\mathbf{W}_\text{key}   \qquad
    \mathbf{V} = \mathbf{X}\mathbf{W}_\text{value}
\label{eq:linear}
\end{equation}
where
$\mathbf{W}_\text{query}$
,
$\mathbf{W}_\text{key}$
and
$\mathbf{W}_\text{value}$
are all matrices of size ${C\times P}$.
In the second and third operations (green and yellow blocks), the \textit{scaled dot-product attention} combines $\mathbf{Q}$, $\mathbf{K}$ and $\mathbf{V}$ as:
\begin{equation}
    \text{Attention}(\mathbf{Q}, \mathbf{K}, \mathbf{V})
    \doteq
    \softmax_\text{over keys} \left(\frac{\mathbf{Q}\mathbf{K}^\text{T}}{\sqrt{P}} \right)\mathbf{V}.
\end{equation}
Noteworthy, the $\frac{\mathbf{Q}\mathbf{K}^\text{T}}{\sqrt{P}}$ block represents the so-called attention matrix, which is used to weigh each element of the \textit{values} tensor with respect to all others based on the relative importance.

The second block comprises two linear layers (orange rectangle in Fig. \ref{fig:bioformers}) that process independently all heads produced as output by the scaled dot-product attention relative to the same sequence element, and project them first to a hidden space and then back to the space $\mathbb{R}^C$.

\section{Material and Methods}
\label{sec:bioformer}
This paragraph introduces Bioformer, a Vision Transformer (ViT) \cite{dosovitskiy2020image} inspired architecture, which significantly reduces the computation complexity for sEMG-based gesture recognition, while reaching an accuracy comparable with the state-of-the-art.
Further, we propose a new pre-training protocol to feed more data to our transformer.
Finally, we provide few details about the experimental setup and the deployment of the networks.
\subsection{Bioformer: Network Topology}
\label{sec:topology}
Our network comprises three modules.
First, the input signal is projected onto a space of dimension $N\times64$ by means of a 1D-convolutional layer.
We use padding = 0 and stride equal to the filter dimension to aggregate non-overlapping windows of the input signal.
Similarly to what is done in ViT for images, the idea is to create a series of $N$ tokens of dimension 64 that encode the input information.
We tested [1, 5, 10, 20, 30] for the filter dimension.
Note that the higher is the dimension, the smaller is the number $N$ of produced tokens.
Therefore, the computation complexity of the following attention blocks' reduces for larger filters.
Compared to standard transformers \cite{vaswani2017attention}, tuning the dimension of this first layer increases the flexibility of the architecture, allowing to trade-off the total number of operations and the accuracy. 

Then, the output of the 1D convolutional layer is processed by the self-attention blocks, described in Sec. \ref{subsec:attention}.
In the rest of the paper, we focus on two variants of the Bioformer architecture, both of which exhibit good accuracy on sEMG-based gesture recognition.
The parameters of the two networks are all identical, except for the number of heads and the number of layers (depth).
The first network comprises one attention layer with eight heads, while the second one consists of two attention layers with two heads each.
These two parameters have been chosen after performing a grid search on depth $\in$ \{1, 2, 3, 4\} and heads $\in$ \{1, 2, 4, 8\}. We chose the architectures with the best trade-off of accuracy vs. parameters.
The hidden space has dimension 128, while each head has a dimension $P$ of 32.
Similarly to \cite{dosovitskiy2020image}, a "class token" is concatenated after the QKV projection step, adding one sample to the sequence length ($(N+1) \times 64$). 
The outputs corresponding to the class token are used to produce the network prediction. Having a dedicated token for prediction, instead of simply using the last (or first) input token, has been shown to yield higher accuracy in \cite{dosovitskiy2020image}. Intuitively, this solution gives higher flexibility to the class token output, which can learn to "pay attention" to relevant elements in the input sequence from the point of view of the classification. 
%
%

The lower section of Fig. \ref{fig:bioformers} summarizes these two network architectures.
\subsection{Bioformer: Training}
\label{sec:training}
The standard training for sEMG gesture recognition, regardless of the employed dataset, is subject-specific, given that the movements and muscle contractions associated with different gestures can differ significantly from one subject to another \cite{zanghieri2019robust, tsinganos2019improved}.
On the other hand, it is known that performing a pre-training step on data similar to the ones used for the final training is highly beneficial for DL models, and in particular for Transformers \cite{devlin2018bert}. For instance, typically, many state-of-the-art image recognition networks do not randomly initialize their weights, but go through a first training on the Imagenet dataset before fine-tuning on their target dataset, especially if the latter is small.
The Imagenet pre-training allows the network to start fine-tuning from a set of well-initialized weights, which can already extract meaningful features for generic images.
Similar pre-training + fine-tuning protocols are also key elements of most successful transformer models, e.g., in NLP \cite{devlin2018bert, brown2020language}. 
Noteworthy, pre-training not only helps to speed up the fine-tuning convergence but also yields higher final accuracy. 

Based on these assumptions, this work introduces a new two-step training procedure for sEMG-based gesture recognition.
%
%
Compared to the standard approach, we first perform an inter-subjects pre-training, in which data relative to all subjects available in the training dataset are employed. Then, we proceed with subject-specific fine-tuning, which is common to all state-of-the-art approaches.
Despite the task being strictly subject-dependent, one can intuitively imagine that the sEMG signal features that are useful for gesture classification should be similar for all patients. Indeed, using this protocol, we observe that feeding more data to the network during pre-training is beneficial for accuracy. 
To clarify better our proposed protocol, we report below the training procedure that we use for subject 1 of the 10-subjects Ninapro DB6 training dataset.
First, we train the network for 100 epochs with data coming from patients 2-10, excluding subject 1, on which we want to test the final model.
In this step, the model adjusts the weights to extract general features that classify the gestures of the other nine patients as accurately as possible.
Then, we perform 20 epochs of fine-tuning using only the training data of subject 1. During this fine-tuning, the recording sessions of the patients are separated between train and test set, following the classical sequential training protocol used by other state-of-the-art approaches for this task, which mimics a real scenario, using sessions 1-5 for training and 6-10 for testing: sessions 1-5 are used, with their golden labels, for training. Then, the trained model is deployed and used to predict gestures belonging to the remaining sessions. 

For the pre-training step, we use Adam optimizer with a linear warmup of the learning rate from 1e-7 to 5e-4; for the fine-tuning step, a fixed learning rate of 1e-4 is used, with a reduction of 10$\times$ after 10 epochs.

\subsection{Experimental Setup \& Dataset}
\label{sec:setup}
To validate our new architectures, we employ the public sEMG-based hand gesture recognition dataset called Non-Invasive Adaptive hand Prosthetics Database 6 (NinaPro DB6)~\cite{PalermoNinaProDB6}, which has been explicitly realized to investigate the degradation of sEMG-based hand gesture recognition accuracy over time. 
The dataset includes 10 non-amputee subjects (3 females, 7 males, average age $27 \pm 6$ years), who have been asked to undergo 10 gathering sessions.
The 10 sessions are distributed over 5 days, one in the morning, one in the afternoon, each including 12 repetitions of the gestures for each patient.
The gestures considered include the rest position and seven grasps, covering hand movements typically done during daily activities. 
Each grasp repetition lasts approximately~\SI{6}{s}, followed by~\SI{2}{s} of rest. 
The array of sensors is composed of 14 Delsys Trigno sEMG Wireless electrodes, placed on the high half of the forearm, simulating the amputation of the lower half of the forearm.
Each sensor gathers the data at a sampling rate of 2~kHz. 
The dataset is divided into windows of 150~ms (i.e., 300 samples) with a slide between them of 15~ms.

For training and validating the model using floating-point (fp32) arithmetic, we employed Python3.7 together with Pytorch1.8.1.
We then perform few epochs of quantization aware training (QAT) to shift from fp32 to integer (int8) arithmetic. We follow the steps described in I-BERT \cite{kim2021bert} to replace the floating-point operators that compose MHSA layers with their int8 counterparts.
Finally, we deployed the resulting quantized models on the GAP8 MCU, using the optimized kernels described in \cite{9524173}. 
GAP8 is a commercial MCU from GreenWaves \cite{gap8}, comprising a controller unit called Fabric Controller (FC), composed of a single RISC-V core, which manages the peripherals and orchestrates the program execution, and a cluster of 8 identical RISC-V cores (with a shared 64 kB scratchpad memory) which can be activated to execute and accelerate intensive workloads. A 512kB main memory is shared between FC and cluster.

\begin{figure}[t]
  \centering
\includegraphics[width=1.1\linewidth]{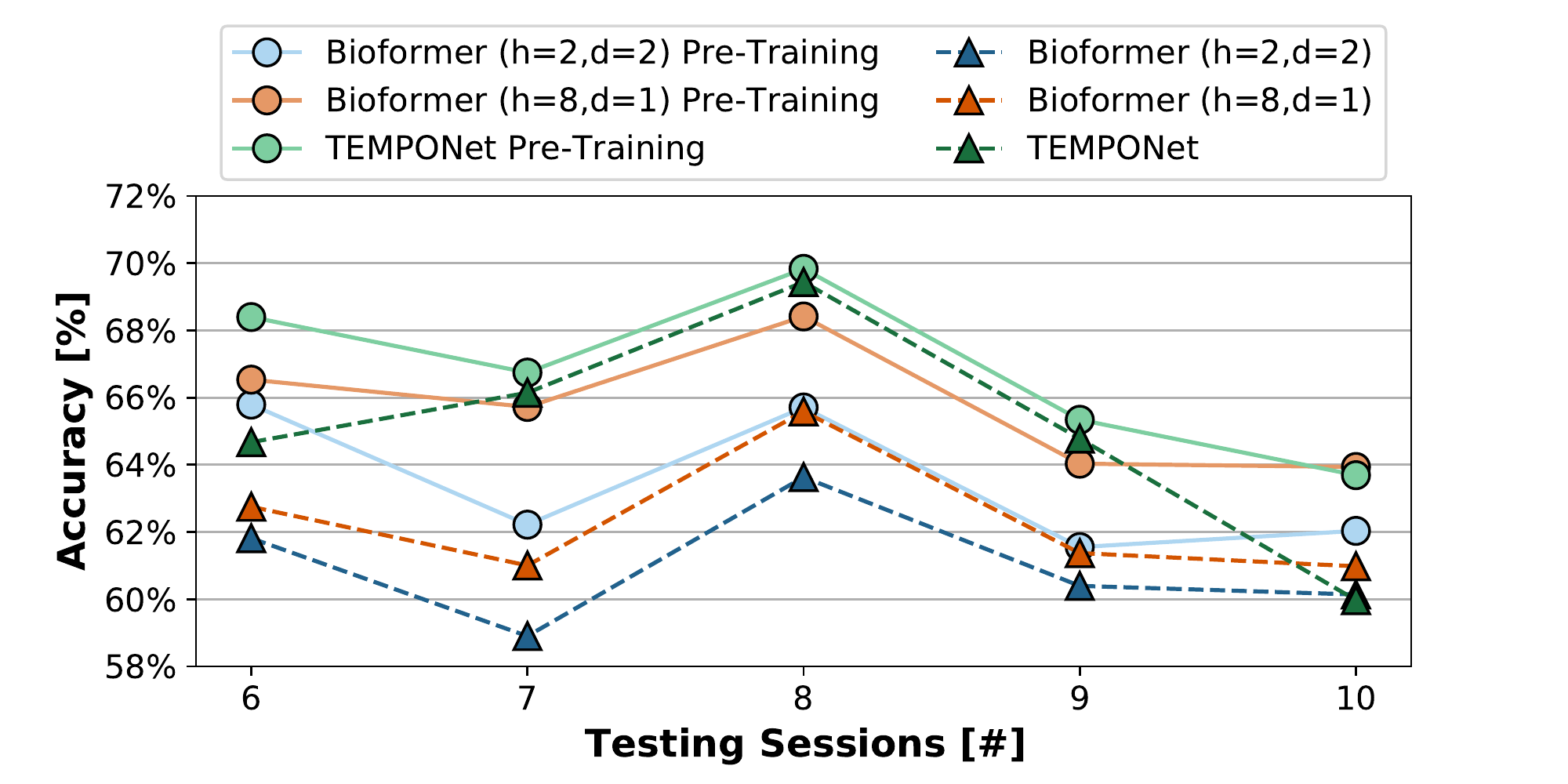}
  \caption{Performance variation on the different testing sessions.}
  \label{fig:training_sessions}
  \vspace{-0.3cm}
\end{figure}
\section{Experimental Results}
\label{sec:experimental_results}
In this section, we first demonstrate the performance of our networks on the Ninapro DB6.
Then, we perform an ablation study to demonstrate i) the pre-training impact on the Bioformer (h = 8, d = 1) and ii) the influence of the filter dimension of the initial convolutional layer.
Finally, we discuss the complexity of our architectures and their latency and energy when deployed on the GAP8 MCU.

\subsection{Ninapro DB6 benchmark}
%
%
Fig. \ref{fig:training_sessions} reports the accuracy of our two Bioformers and of the state-of-the-art TEMPONet \cite{zanghieri2019robust}. Each point corresponds to one of the five testing sessions, and the reported accuracy is the average across patients. Higher session numbers correspond to tests farther in time from the training period.
Compared to the reference TCN, our Bioformers achieve slightly lower accuracy both with and without pre-training. 
Bioformers without pre-training achieve a 2.7\%-3.9\% lower accuracy on average. However, the accuracy difference w.r.t. TEMPONet decreases for sessions that are farther in time from the training and therefore more dissimilar from it.  In particular, the h=8, d=1 Bioformer outperforms TEMPONet on testing session 10 (+ 0.48\%). This result suggests that thanks to the reduced number of parameters (reported below) our architecture is more prone to well generalize on more dissimilar data, a key factor for a task where the data show high variability over time.
Noteworthy, applying the pre-training is beneficial both for the proposed Bioformers and for TEMPONet. However, the accuracy difference between the two types of models decreases, confirming the superior capability of Transformer-based architectures to take benefit from pre-training with large amounts of data.
On the different sessions, we observe an average gain of 3.39\%, 2.48\%, and 1.80\% for Bioformer (h=8, d=1), Bioformer (h=2, d=2), and TEMPONet, respectively.

Overall, our best architecture (i.e., the one with 8 heads) achieves an average 65.73\% accuracy, which is 0.73\% better than the previous state-of-the-art TEMPONet, and 1.07\% lower than the new pre-trained TEMPONet. 
\subsection{Ablation Study: pre-training \& Patch Dimension}
\begin{figure}[t]
  \centering
\includegraphics[width=1.1\linewidth]{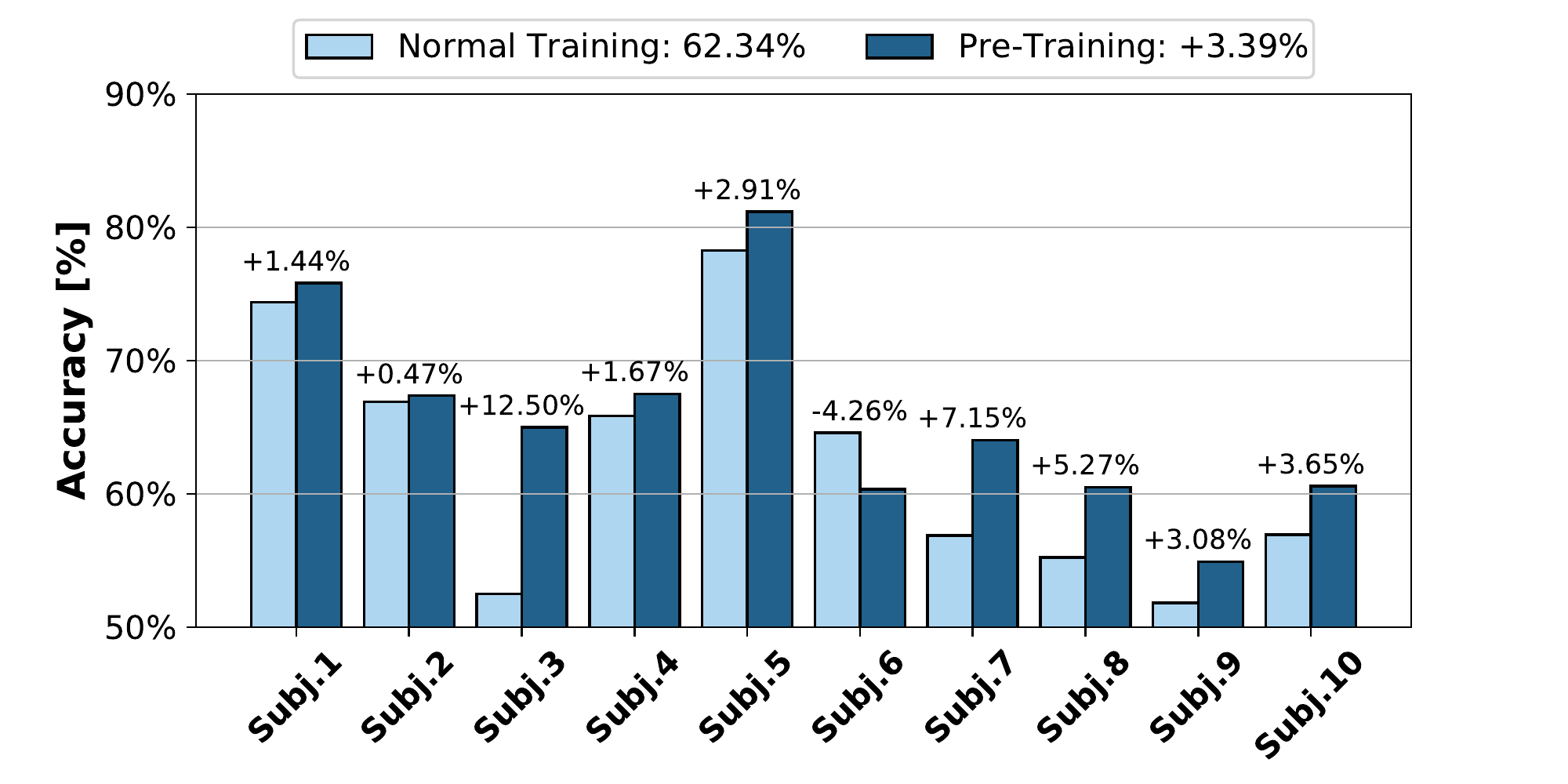}
  \caption{Accuracy per subject with intra- and inter- patient training data.}
  \label{fig:pretraining}
  \vspace{-0.6cm}
\end{figure}
\begin{figure}[t]
  \centering
\includegraphics[width=1.1\linewidth]{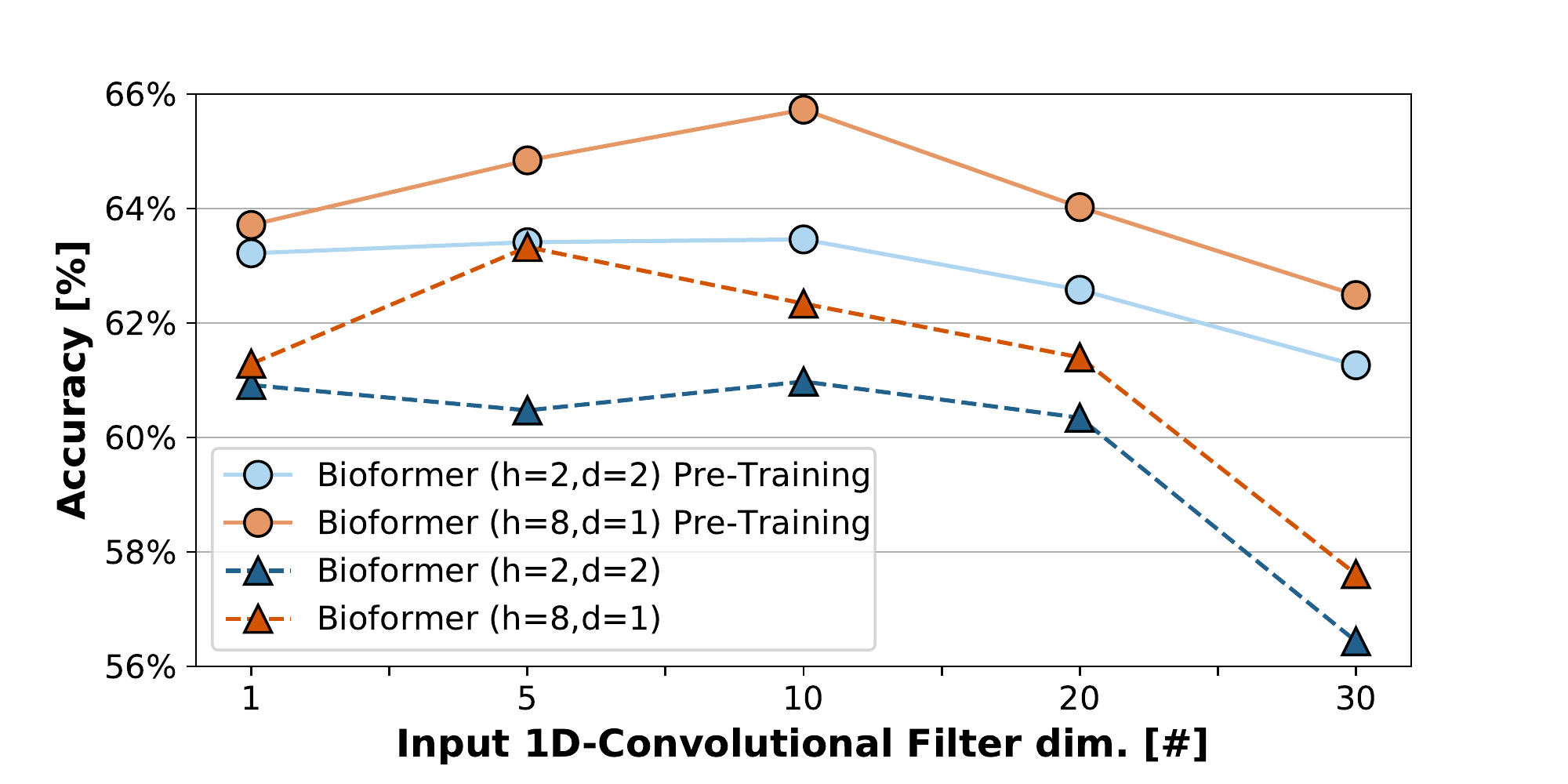}
  \caption{Performance using [1,30] filter dimensions for the front-end convolutional layer. Increasing filter dimension reduces both the number of parameters and the number of operations.}\label{fig:patch_dim}
  \vspace{-0.4cm}
\end{figure}
In this paragraph, we detail i) the benefit of applying our new training approach and ii) the impact of the filter dimension of the initial 1D convolutional layer in Bioformers.

Fig. \ref{fig:pretraining} details the performance change between standard training and our two-steps training for each subject.
We can notice that the most significant advantages are obtained for subjects that present the lower accuracy before pre-training.
On subjects whose starting accuracy is lower than 60\%, the average accuracy improvement is 6.33\%, while on the other five subjects, it is just 0.45\%, leading to an overall average improvement of 3.39\%. Solely the Subj.6's accuracy get worse with our new training. This could be caused by the lower learning rate used in the subject-specific fine-tuning that does not allow the network to converge to the global minimum. 

In Fig. \ref{fig:patch_dim}, we show the impact of the filter dimension of the first convolutional layer. Note that 1D convolution is always applied in a non-overlapping fashion in our networks. Therefore, a wider filter implies a smaller input signal for the attention block. Each solid line represents a Bioformer on which we applied the two-step training (pre-training and fine-tuning), whereas the dashed lines correspond to networks trained with the standard procedure.
For most models, a filter dimension equal to 10 results in the best accuracy, despite its lower complexity compared to 1 and 5 (the resulting input sequence length is 30 instead of 60 and 300 for filter sizes 5 and 1\footnote{When a filter size of 1 is applied, the 1D-convolutional layer becomes a fully-connected embedding layer.}, respectively).
Furthermore, despite the resulting lower accuracy, increasing the filter dimension beyond 10 can be beneficial from the deployment point of view, given the reduction in the algorithm's complexity, whose number of operations depends almost linearly on the sequence length. 
For instance, changing the dimension from 10 to 20, employing the Bioformer (h=8, d=1) only causes a drop of 1.70\% of accuracy, while reducing the total number of operations by a factor 1.93$\times$, and the energy by 2$\times$, with a potentially very significant impact on the battery life of the device executing the inference.
\subsection{Deployment on GAP8}
\begin{figure}[t]
\centering
\begin{subfigure}{0.5\textwidth}
\includegraphics[width=1.1\linewidth, trim=0cm 0cm 0cm 0cm, clip]{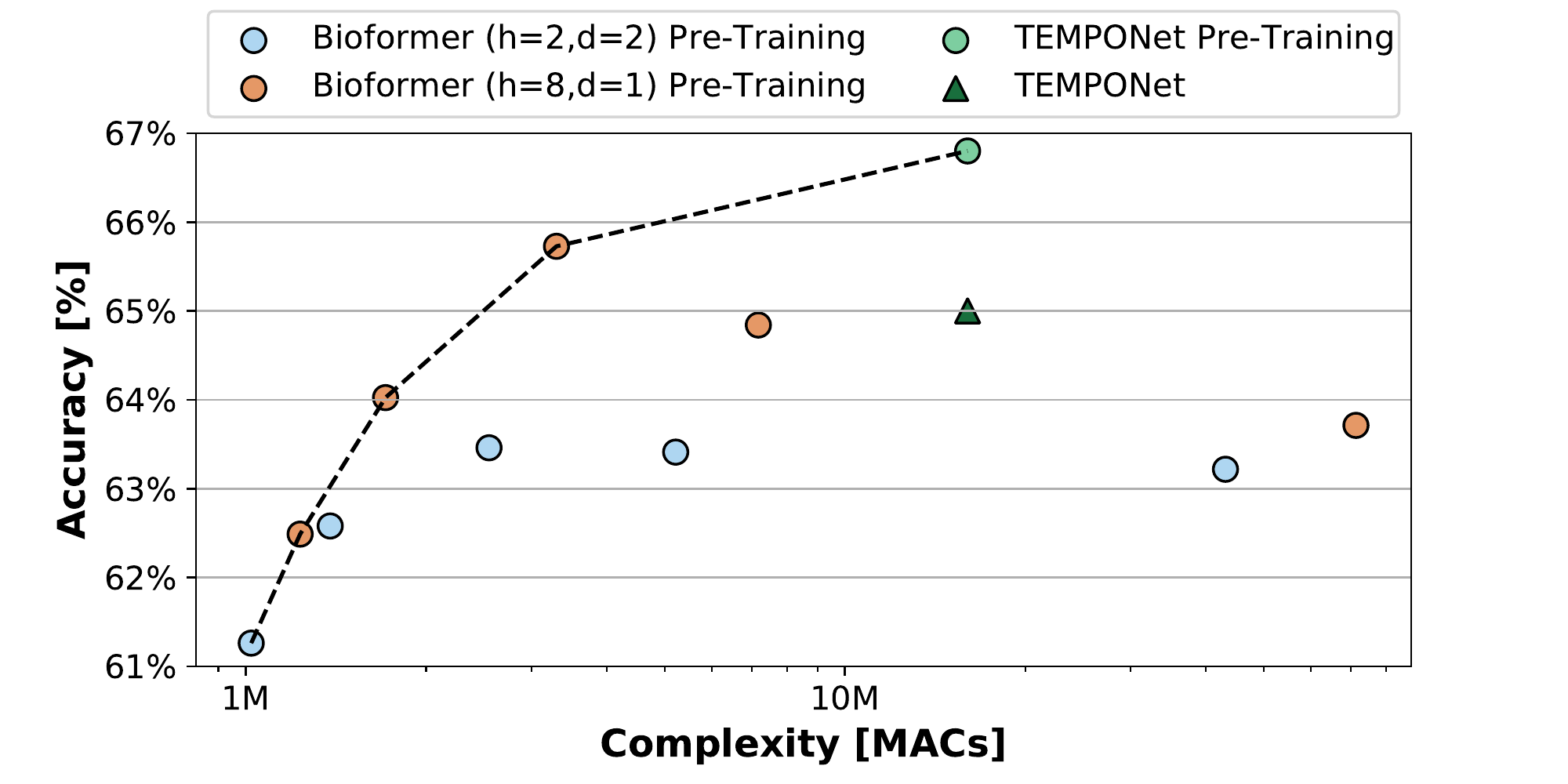}
\subcaption[]{Accuracy vs parameters.}
\end{subfigure}\hspace{\textwidth}%
\begin{subfigure}{0.5\textwidth}
\includegraphics[width=1.1\linewidth, trim=0cm 0cm 0cm 0cm, clip]{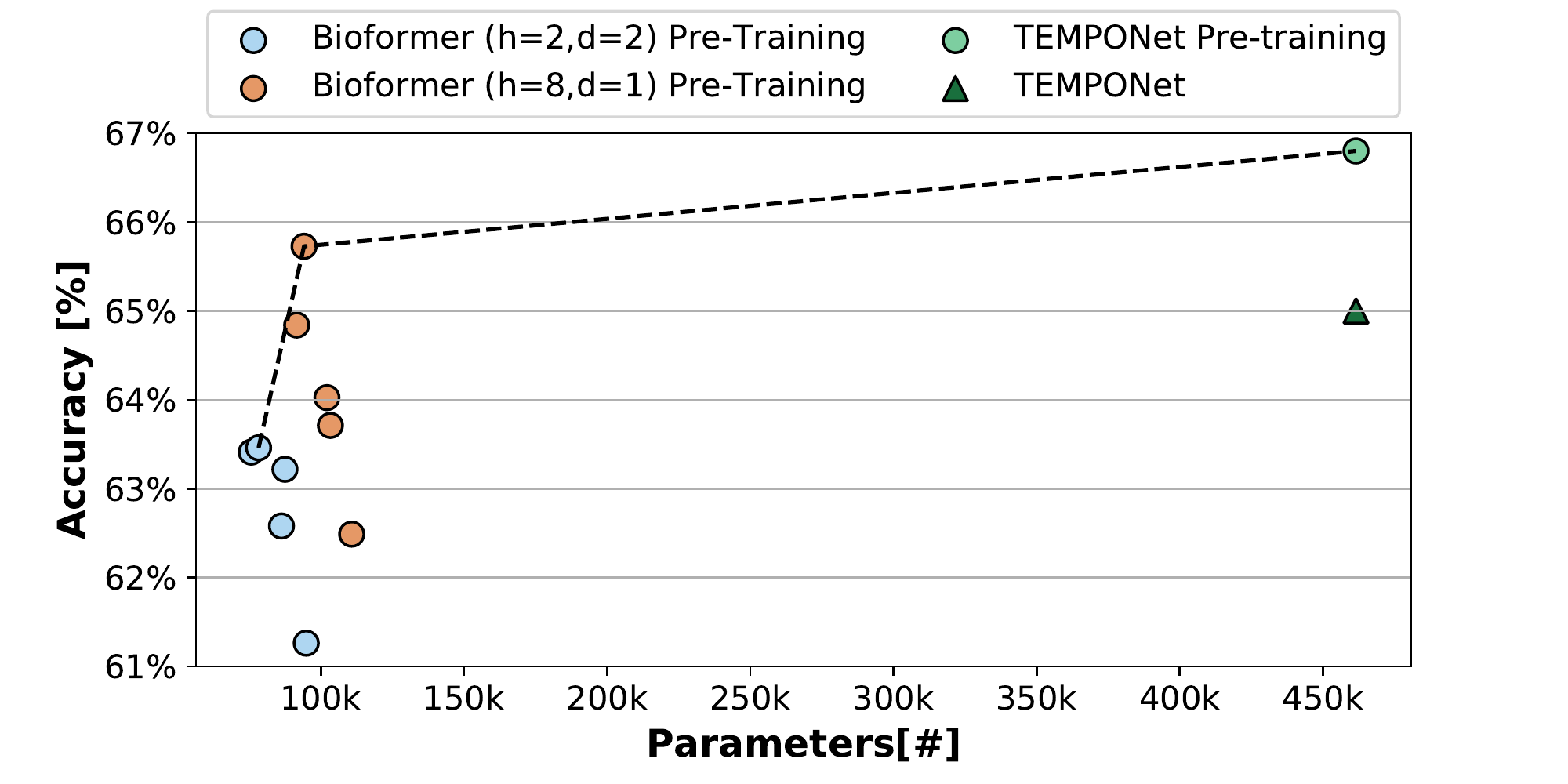}
  \subcaption{Accuracy vs MAC operations.}
\end{subfigure}
\caption{Pareto spaces}
  \label{fig:pareto}
  \vspace{-0.4cm}
\end{figure}
Fig. \ref{fig:pareto} shows different Bioformer architectures as well as TEMPONet in the N. of Operations versus accuracy and N. of parameters vs accuracy planes.
While the pre-trained TEMPONet reaches the highest accuracy, all other Pareto points are populated by Bioformers.
The different points plotted for the same Bioformer refer to different filter sizes of the initial 1D Convolutional layer.
In the complexity versus accuracy space, we identified two key architectures of our Bioformers.
Our most accurate model (h=8, d=1, filter = 10) outperforms the state-of-the-art TEMPONet and is only 1.07\% less accurate than the pre-trained TEMPONet, but showing an impressive 4.9$\times$ operations reduction.
The lightest Bioformer (h=2, d=2, filter = 10) on the Pareto frontier, instead, reduces the required number of operations of an additional factor 3.3$\times$ (16.17$\times$ lower than TEMPONet), at the cost of an additional 4.47\% accuracy drop.
Furthermore, in the lowermost graph, we can observe that all our models have a comparable number of parameters. In fact, the modification of the filter dimension of the 1D convolution only impacts the number of parameters of the first layer and of the linear layers contained in MHSA blocks, with a limited impact on the total model size.

\begin{table}[t]
\centering
\caption{Performance of the quantized Pareto architectures on the GAP8 MCU. Bio1 corresponds to Bioformer (h=8, d=1), Bio2 to Bioformer (h=2, d=2).\\Abbreviations: Lat.: latency, E.: energy, Q.Acc.: quantized accuracy.}
\label{tab:network}
\begin{tabular}{l|lllll} 
Network  & Memory & MMAC & Lat.[ms] & E.[mJ] & Q. Acc.   \\ \hline
\multicolumn{6}{l}{MCU: GAP8, 100 MHz @ 1V, 51 mW} \\ \hline
Bio1, wind=30 & 110.8 kB & 1.2 & 1.03  & 0.052 & 61.09\%\\
Bio1, wind=20 & 102.1 kB & 1.7 & 1.37 & 0.070 & 63.14\%\\
Bio1, wind=10 & 94.2 kB & 3.3 & 2.72 & 0.139 & 64.69\%\\
Bio2, wind=30 & 92.2 kB & 1.0 & 1.55 & 0.079 & 60.19\%\\
Bio2, wind=10 & 78.3 kB & 2.5 & 4.82 & 0.246 & 62.43\%\\ \hline
TEMPONet \cite{zanghieri2019robust} & 461 kB & 16.0 & 21.82 & 1.11 & 61.00\%\\\hline                           
\end{tabular}
  \vspace{-0.4cm}
\end{table}
The results of deploying some of these Pareto architectures on GAP8 are shown in Table \ref{tab:network}. The average power consumption while using the 8-cores cluster to execute the Bioformer inference is 51 mW @ 100 MHz. For TEMPONet, we report its statistic in an identical setup, to allow for a fair comparison between the two models.
Note that the accuracy of these models, as reported in Table \ref{tab:network}, is the one obtained after the quantization-aware fine-tuning.

After quantization, our most accurate model yet achieves 64.69\% accuracy, consuming an impressively lower 8.0$\times$ energy compared to TEMPONet.
Additionally, this model can even fit a smaller MCU since it only requires 94.2 kB.

The Bioformer with the lowest latency further reduces the energy compared to TEMPONet by 17.3$\times$, with an accuracy reduction of only 3.60\% and a comparable memory footprint (110.8 kB).
Overall, considering this last model, a 150 ms window classified every 15 ms costs 52 $\mu$J and has a latency of 1.02 ms, while for the remaining time, the GAP8 SoC only collects data. In this step, we can idle the 8-core cluster accelerator using its embedded hardware synchronization unit \cite{8715266} and therefore reduce the power consumption to the 10 mW consumed by the Fabric Controller.
This yields an average power consumption over time of 12.81 mW. Using a small \SI{1000}{mAh} battery, we can continuously perform sEMG-based gesture recognition for a lifetime of $\sim 257 h$, 4.77$\times$ higher compared to using TEMPONet for this task ($\sim 54 h$ with the same battery \cite{zanghieri2019robust}).
\section{Conclusions}
\label{sec:conclusion}
We have shown that Transformers can achieve close to state-of-the-art performance on sEMG-based gesture recognition while strongly reducing the complexity and the memory footprint required for deployment on edge nodes.
Further, we have introduced a new pre-training procedure that yields up to a 3.39\% accuracy improvement for both Transformer and TCN-based models. 

On Ninapro DB6, our most accurate Bioformer obtains 65.73\% accuracy, better than the previous state-of-the-art accuracy (65.00\% of TEMPONet \cite{zanghieri2019robust}) and only 1.07\% lower than TEMPONet trained with our new protocol.
Simultaneously, this Bioformer reduces the number of MACs and memory footprint compared to TEMPONet by 4.9$\times$.
Deployed on GAP8, it consumes just 0.139 mJ with a latency of 2.72 ms.

\tiny
\bibliographystyle{IEEEtran}

\end{document}